\documentclass{article}      
\usepackage{graphicx}
\usepackage{epstopdf}
\title{Electromagnetic multipoles - theory issues}  

\author{M.M. Giannini\\
Dipartimento di Fisica and INFN, Genova, Italy\\
}

\date{}
\begin{document}

\maketitle  

\begin{abstract}
  Some predictions of the Hypercentral Constituent Quark Model for the helicity amplitudes are
discussed and compared with data and with the recent analysis of the Mainz group; the role
of the pion cloud contribution in explaining the major part of the missing strength at
low $Q^2$ is emphasized.
\end{abstract}

\section{The hypercentral Constituent Quark Model}

In the hypercentral Constituent Quark Model (hCQM) one introduces the
hyperspherical coordinates, which are obtained from the standard Jacobi
coordinates $\vec{\rho}$ and $\vec{\lambda}$
substituting the absolute values $\rho$ and $\lambda$ by
\begin{equation}
x=\sqrt{{\vec{\rho}}^2+{\vec{\lambda}}^2} ~~,~~ \quad
\xi=arctg(\frac{{\rho}}{{\lambda}}),
\end{equation}                                       
\noindent where $x$ is the hyperradius and $\xi$ the hyperangle.
The potential for the three quark system, $V$, is assumed to depend on the
hyperradius $x$ only, that is to be hypercentral. It can be considered as
a two-body interaction in the hypercentral approximation, which has been
shown to be valid specially for the lower energy states \cite{hca}. It
can also be viewed as a true three-body potential; actually the fundamental gluon
interactions, predicted by QCD, lead to three-quark mechanisms. The situation is
similar to the flux tube models, where two-body ($\Delta$-shaped) and three-body
($Y$-shaped) interactions are considered.

For a hypercentral potential, in the three-quark wave function one can
factor out the angular and hyperangular parts, which are given by the
known hyperspherical harmonics \cite{bf} and the Schr\"{o}dinger
equation is reduced to a single equation for the hypercentral wave
function. Such hypercentral equation can be solved analytically at least in two
cases, that is for the h.o. potential and the hypercoulomb one. The
two-body h.o. potential turns out to be exactly
hypercentral, since $\sum_{i<j}~\frac{1}{2}~k~(\vec{r_i}
- \vec{r_j})^2~=~\frac{3}{2}~k~x^2~$. The $SU(6)$ states in the h.o. model
are too degenerate with respect to the observed spectrum. The
'hypercoulomb' potential \cite{hca,br} $V_{hyc}(x)= -\frac{\tau}{x}$
is not confining, however it leads to a power-law behaviour of
the proton form factor and of all the transition form factors
\cite{sig} and it has a perfect degeneracy between the first $0^+$
excitated state and the first $1^-$ states. The former can
be identified with the Roper resonance and the latter with the negative
parity resonances. This degeneracy seems to be in agreement with phenomenology but such
feature cannot be reproduced in models with only two-body forces, since the excited $L = 0$
state, having one more node, lies above the $L=1$ state.

In the hCQM \cite{pl} the confining hypercentral potential is assumed to
be of the form
\begin{equation}\label{eq:pot}
V(x)= -\frac{\tau}{x}~+~\alpha x,
\end{equation}                                                             
A standard hyperfine interaction
\cite{ik}, treated as a perturbation, is added in order to describe the
splittings within the
$SU(6)$ multiplets. The non strange spectrum is described with
$\tau~=~4.59$ and $\alpha~=~1.61~fm^{-2}$ and the standard strength of the
hyperfine interaction needed for the $N-\Delta$ mass difference
\cite{ik}. The model, keeping fixed these three parameters, has been
applied in order to calculate, that is predict, various quantities of
interest, namely the photocouplings \cite{aie}, the transition helicity
amplitudes \cite{aie2}, the elastic nucleon form factors \cite{mds} and
the ratio between the electric and magnetic form factors \cite{rap}. In
the following the results of this model for the transition helicity
amplitudes will be discussed.

The model has been modified in two respects in order to improve the
description of the spectrum. First, isospin dependent terms have been
added to the spin-spin ones \cite{iso}; the second modification is that to use the
correct relativistic kinetic energy \cite{rel}. The resulting spectrum is considerably
improved, in particular the correct ordering of the Roper resonance and the negative
parity states is achieved.

\section{The helicity amplitudes}

The electromagnetic transition amplitudes, 
$A_{1/2}$ and $A_{3/2}$, are defined as the matrix elements of
the  transverse electromagnetic interaction, $H_{e.m.}^t$, between the nucleon, 
$N$, and the resonance, $B$, states:
\begin{equation}
\begin{array}{rcl}
A_{1/2}&=& \langle B, J', J'_{z}=\frac{1}{2}\ | H_{em}| N, J~=~
\frac{1}{2}, J_{z}= -\frac{1}{2}\
\rangle\\
& & \\
A_{3/2}&=& \langle B, J', J'_{z}=\frac{3}{2}\ | H_{em}| N, J~=~
\frac{1}{2}, J_{z}= \frac{1}{2}\
\rangle\\\end{array}
\end{equation}

The baryon states are obtained using the hCQM: 
\begin{equation}
V_{3q}÷=÷-\frac{\tau}{x}÷+÷\alpha÷x÷+H_{hyp}\label{eq:hyp}
\end{equation}
with the parameters fixed in the previous section.

The transverse transition operator is assumed to be
\begin{equation}\label{eq:htm}
H^t_{em}~=~-~\sum_{i=1}^3~\left[\frac{e_j}{2m_j}~(\vec{p_j} \cdot \vec{A_j}~+
~\vec{A_j} \cdot \vec{p_j})~+~2 \mu_j~\vec{s_j} \cdot (\vec{\nabla} 
\times \vec{A_j})\right]~~,
\end{equation}
where spin-orbit and higher order corrections are neglected 
\cite{cko,ki}. In 
Eq. \ref{eq:htm} $~~m_j$, $e_j$, $\vec{s_j}$ , $\vec{p_j}$ and $\mu_j~=~\frac{ge_j}{2m_j}$
denote the mass, the electric charge, the spin, the momentum and the magnetic 
moment of the j-th quark, respectively, and $\vec{A_j}~=~\vec{A_j}(\vec{r_j})$
is the photon field.

The proton photocouplings of the hCQM \cite{aie} have the same overall behaviour of
other CQM, probably because all models
have the same SU(6) structure in common. In
many cases the strength is underestimated and this is a problem for all CQMs.

Taking into account the $Q^2-$behaviour of the transition matrix elements, one can
calculate the hCQM helicity amplitudes in the Breit frame \cite{aie2}. The hCQM
results for the S11(1535) resonance
\cite{aie2} are given in Fig. 1. The agreement is remarkable, the more so since the
hCQM curve has been published three years in advance with respect to the recent TJNAF
data \cite{dytman}.  In general the $Q^2$ behaviour of the helicity amplitudes is
reproduced, except for discrepancies at small $Q^2$, especially in the
$A_{3/2}$ amplitudes.  These discrepancies could be ascribed either to the
non-relativistic character of the model or to the lack of explicit quark-antiquark
configurations, which may be important at low $Q^{2}$ .  However, the kinematical
relativistic corrections at the level of boosting the nucleon  and the resonances states to
a common frame are not  responsible for these discrepancies,  as it has been demonstrated
in Ref.\cite{mds2}.

\begin{figure}[ht]
\begin{center}
\includegraphics[width=5in] {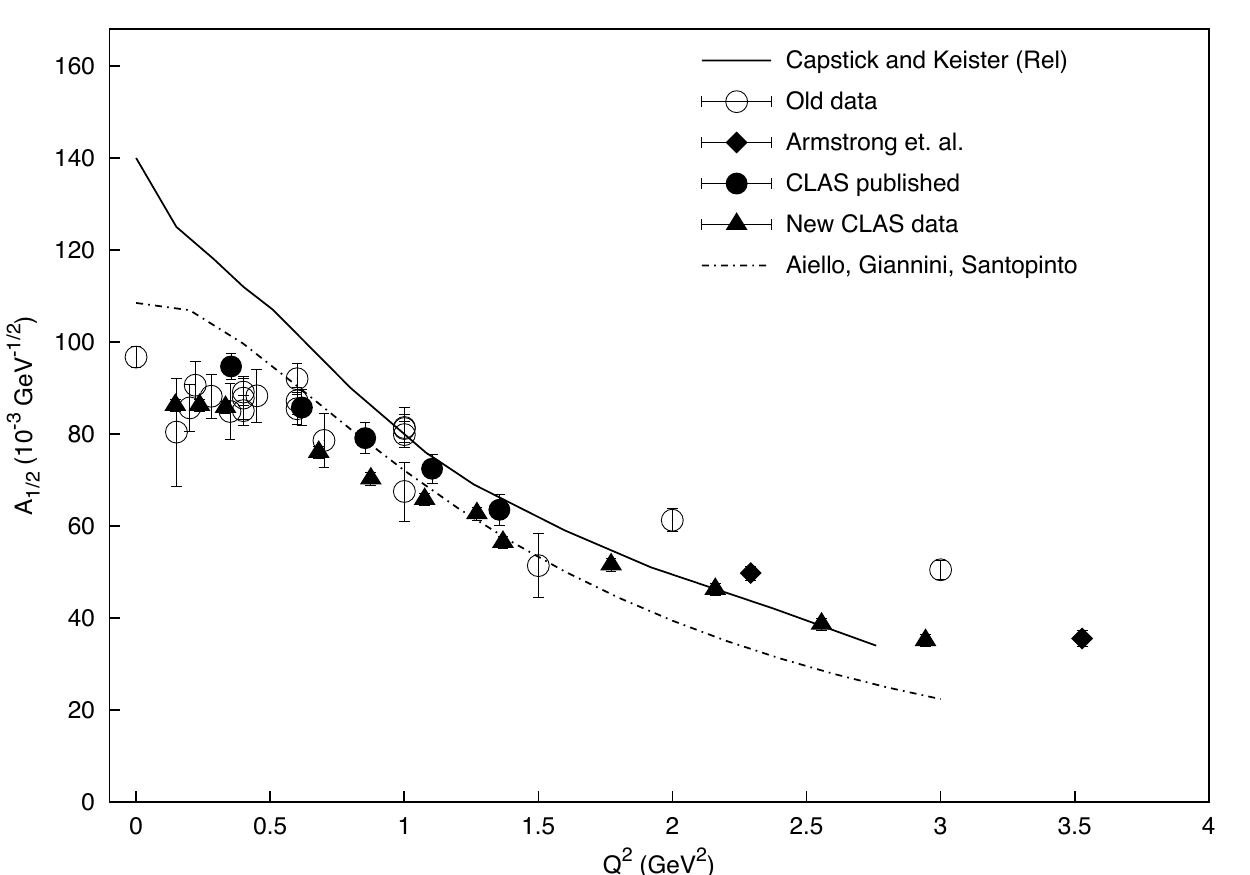}

\caption{Comparison between the experimental data \cite{burk,dytman} for the
helicity amplitude $A^p_{1/2}$ of the $S_{11}(1535)$
resonance and the calculations with the hCQM, lower curve \cite{aie2} 
also compared with the Capstick and Keister result, upper curve \cite{capstick}. 
}
\end{center}
\end{figure}

Keeping fixed the parameters, the hCQM has also been applied to the calculation of
the longitudinal helicity amplitudes \cite{long}. An interesting feature is that
many amplitudes vanish in the $SU(6)$ limit, therefore a detailed study of the
longitudinal strength may be a good test of the $SU(6)$ breaking mechanisms.

It should be mentioned that the r.m.s. radius of the proton corresponding to the
parameters of Eq.\ref{eq:hyp} is $0.48~fm$, which is just the value fitted in
\cite{cko} to the $D13$ photocoupling. Therefore the missing strength at low $Q^2$ 
can be ascribed to the outer region of the nucleon, where the lack of
quark-antiquark effects are probably important. This
view is enforced by a recent analysis \cite{lt_brag,ts03hel}, which compares
the results of the hCQM for the helicity amplitudes and the calculation of the pion
cloud contributions performed with the dynamical model of the Mainz Group. As an example,
the $A_{3/2}$ for the $N-\Delta$ transition is shown in Fig. 2. The pion cloud turns out to
be important at low
$Q^2$ and diminishes strongly up to $3 ÷GeV^2$; it accounts for the major part of the
discrepancy between the data and the hCQM results. Particularly important is the
longitudinal $\Delta$ transition, where the very small hCQM values are compensated by the
dominant pion contribution (see \cite{lt_brag}).

\begin{figure}[ht]
\begin{center}
\includegraphics[width=5in] {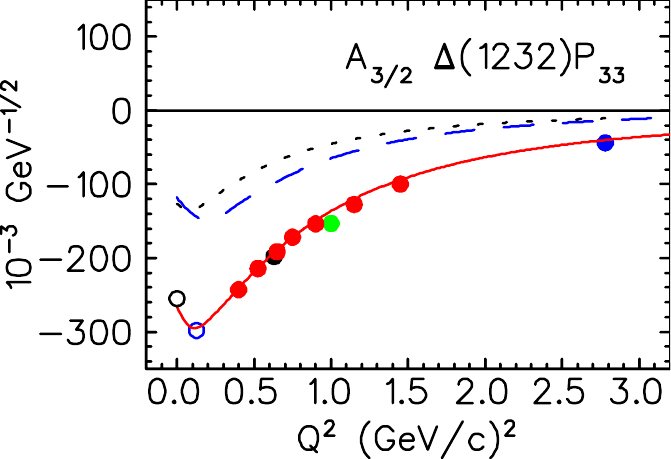}

\caption{ The $Q^2$ dependence
of the $N \rightarrow \Delta$ $A_{3/2}$ helicity amplitude. The solid curve is the result
of the superglobal fit with MAID \cite{Maid}, the data points at finite $Q^2$ are obtained
with single-Q$^2$ fits \cite{ts03hel}. The dashed and dotted curves are, respectively, the
predictions of the hyperspherical constituent quark model
\cite{aie2} and the pion cloud contributions calculated with DMT by
the Mainz group \cite{DMT}. At $Q^2=0$ the photon coupling from PDG is shown
\cite{PDG02}. }
\end{center}
\end{figure}

%
%
%
%

\end{document}